\documentclass{emulateapj}
\def\stacksymbols #1#2#3#4{\def\theguybelow{#2}
        \def\verticalposition{\lower#3pt}
        \def\spacingwithinsymbol{\baselineskip0pt\lineskip#4pt}
        \mathrel{\mathpalette\intermediary#1}}
\def\intermediary #1#2{\verticalposition\vbox{\spacingwithinsymbol
        \everycr={}\tabskip0pt
        \halign{$\mathsurround0pt#1\hfil##\hfil$\crcr#2\crcr
                \theguybelow\crcr}}}
\def\lta{\stacksymbols{<}{\sim}{2.5}{.2}}
\def\gta{\stacksymbols{>}{\sim}{3}{.5}}

\shorttitle{Cooling Cavities}
\shortauthors{Brighenti, Mathews \& Temi}

\begin{document}

\title{
Hot gaseous atmospheres in galaxy groups and clusters 
are both heated and cooled by X-ray cavities
}

\author{
Fabrizio Brighenti\altaffilmark{1,2}, 
William G. Mathews\altaffilmark{1}, 
Pasquale Temi\altaffilmark{3}
}

\altaffiltext{1}{University of California Observatories/Lick
  Observatory,
Department of Astronomy and Astrophysics,
University of California, Santa Cruz, CA 95064
(mathews@ucolick.org).}
\altaffiltext{2}{Dipartimento di Astronomia,
Universit\`a di Bologna, via Ranzani 1, Bologna 40127, Italy
(fabrizio.brighenti@unibo.it).}
\altaffiltext{3}{Astrophysics Branch, NASA/Ames Research Center, 
MS 245-6,
Moffett Field, CA 94035 (pasquale.temi@nasa.gov).}

\begin{abstract}
Expanding X-ray cavities observed in hot gas atmospheres of many
galaxy groups and clusters generate shock waves and turbulence that
are primary heating mechanisms 
required to avoid uninhibited radiatively cooling flows
which are not observed.
However, we show here that the evolution of buoyant cavities
also stimulates 
radiative cooling of observable masses of low-temperature gas.
During their early evolution, 
radiative cooling occurs in the wakes of buoyant cavities 
in two locations: 
in thin radial filaments parallel to the buoyant velocity 
and more broadly in gas compressed beneath rising cavities.
Radiation from these sustained compressions removes 
entropy from the hot gas.
Gas experiencing the largest entropy loss cools first,
followed by gas with progressively less entropy loss. 
Most cooling occurs at late times, $\sim 10^8-10^9$ yrs, 
long after the  
X-ray cavities have disrupted and are impossible to detect.
During these late times, slightly denser low entropy gas 
sinks slowly toward the centers of 
the hot atmospheres where it cools intermittently, forming clouds 
near the cluster center.
Single cavities of energy $10^{57}-10^{58}$ ergs
in the atmosphere of the NGC 5044 group
create $10^8 - 10^9$ $M_{\odot}$ of cooled gas,
exceeding the mass of extended molecular gas currently observed 
in that group.
The cooled gas clouds we compute share many attributes with
molecular clouds recently observed in NGC 5044 with ALMA:
self-gravitationally unbound, dust-free, quasi-randomly
distributed within a few kpc around the group center.
\end{abstract}

\section{Introduction}

Cavities are visible in the hot atmospheres of 
about 30 percent of all X-ray bright galaxy groups and clusters,  
but this percentage rises sharply with decreasing 
radiative cooling time of the central gas  
(e.g. Panagoulia et al. 2014).
It is generally thought that 
powerful cavity-producing jets are released by
cluster-centered massive black holes soon after 
accreting a small amount of cooled cluster gas.
By feeding energy back into the cluster gas,
the black hole increases the entropy
and cooling time of nearby (and distant) gas 
in the surrounding atmosphere, 
maintaining the accretion rate at an acceptably low level.
Weak shock waves produced by jets and their expanding 
cavities propagate far into the hot atmosphere, 
dissipating their energy. Cavity buoyant motion generates subsonic
turbulence which can also contribute to gas heating (Dennis \&
Chandran 2005).
By measuring the PV work done on the cluster gas by
the cavities,
it is possible to determine a lower limit to the 
rate that the atmospheres gain energy, 
although the total power delivered may greatly 
exceed this lower limit (Mathews \& Guo 2011).

While cavities are likely to be a primary
heating mechanism required to offset radiative cooling 
in hot group/cluster atmospheres,
our goal here is to demonstrate that the mere 
presence of cavities,
regardless of the mechanism that forms them,
also stimulates profound inhomogeneous cooling in 
the hot gas over long times 
(see also Brighenti \& Mathews 2002).
We do not consider the entire AGN feedback process in this paper 
in which cooling flows are suppressed
by recurrent bubble heating (see,
for example Brighenti \& Mathews 2002, 2003, Dalla Vecchia et al.,
2004, Mathews 2009),
but instead limit our computations to study the
influence of single X-ray cavities on the surrounding gas. 
Much is known from recent computational studies
about the dynamics of cavity formation and their 
subsequent evolution 
(e.g. Bruggen  2003;
Jones \& De Young 2005;
Reynolds et al. 2005;
Gardini 2007;
Ruszkowski et al. 2007:
Revaz, Combes \& Salome 2008;
Mathews \& Brighenti 2008;
Bruggen, Scannapieco \& Heinz, 2009;
Dong \& Stone 2009;
O'Neill, De Young \& Jones 2009;
Guo \&Mathews 2011), 
but few of these studies examine 
localized radiative cooling 
stimulated by cavity evolution.
It is widely known that buoyantly rising 
cavities cause outflowing streams or filaments 
of (low entropy) cluster gas to rise up behind them. 
Revaz, Combes \& Salome (2008) demonstrate
that some of this rising gas cools radiatively,
possibly explaining  
radial filaments observed in H$\alpha$ 
(e.g. Conselice et al. 2001) as well as 
kinematically related columns of colder molecular gas
(Salome \& Combes 2004;
Ho et al. 2009; Lim et al. 2012).
Recent investigations suggest that spatially extended cooling can
occur in galaxy clusters where the ratio of cooling time to dynamical
time is $t_{cool}/t_{dyn} \la 10$, provided the ICM is 
endowed with non-linear density
perturbations (McCourt et al. 2012, Sharma et al. 2012, Gaspari et
al. 2012). 
Our main objective is to show that the buoyant motion of cavities 
stimulates intermittent cooling in the hot virialized gas in 
galaxies, groups and clusters.
Cavity-cooled gas is  
distributed over kpc scales and lasts $10^8-10^9$ yrs.
No {\it ad hoc} distributed heating or artificial
turbulent velocity field are necessary to form cooling perturbations.
Valentini \& Brighenti (2015) explore AGN
induced cooling process in massive and intermediate elliptical
galaxies and show that it occurs for $t_{cool}/t_{dyn} \la 70$.
 
Molecular CO(2-1) emission was recently discovered 
in the bright group-centered elliptical 
galaxy NGC 5044 
with single-dish IRAM sub-millimeter observations
(see David et al. 2014).  
The molecular gas has 
a total mass $\sim10^8$ $M_{\odot}$. 
This curious and unexpected result 
was confirmed with ALMA observations 
by David et al. (2014)
who determined that much of the CO(2-1) emission was due to 
24 spatially distinct clouds or cloud-aggregations in the 
hot atmosphere within 2 kpc of the group center.
NGC 5044 has a vigorously active
central black hole that feeds accretion energy back into its hot
atmosphere -- this is not the astronomical
environment traditionally thought to create 
or harbor molecular clouds.
The CO clouds are spatially unrelated to extended,
optically absorbing dust patches seen in NGC 5044. 
They appear to be dust-free and self-gravitationally unbound.
This suggests that the clouds may be currently cooling  
directly from the hot group atmosphere in NGC 5044 
in a manner 
similar to the inhomogeneous cooling regions we calculate here.

In the following we review some of the numerical 
difficulties encountered when computing 
radiative cooling in Eulerian hydrodynamic flows. 
Fortunately, these difficulties can be largely overcome
with a subgrid cooling scheme 
and we apply this scheme to cavity dynamics in group/cluster 
atmospheres having multi-keV temperatures.
We find that 
radiative cooling is enhanced not only in 
radial filaments beneath buoyant cavities but also 
across the base of the cavities.
Even more remarkable, we find that additional significant
off-center radiative cooling may persist long after 
the observable lifetimes of the cavities responsible.

\section{Computational Strategy}

We consider solutions of the standard hydrodynamic equations:
\begin{equation}
{ \partial \rho \over \partial t}
+ {\bf \nabla} \cdot \rho {\bf u} = 0
\end{equation}
\begin{equation}
\rho \left( { \partial {\bf u} \over \partial t}
+ ({\bf u} \cdot {\bf \nabla}){\bf u} \right)
= - {\bf \nabla} P 
+ \rho {\bf g}
\end{equation}
and 
\begin{equation}
{\partial e \over \partial t}
+ {\bf \nabla} \cdot {\bf u}e = -P ({\bf \nabla} \cdot {\bf u})
-\left({\rho \over m_p}\right)^2 \Lambda
\end{equation}
where $e=(\gamma -1)P$ is the thermal energy density.
The coefficient $\Lambda(T)$ in the term for 
optically thin radiative cooling is that of 
Sutherland \& Dopita (1993), 
evaluated with solar abundances.

We solve these equations in 2D cylindrical coordinates 
using a staggered ZEUS code (Stone \& Norman 1992). 
2D hydro allows us to consider very high 
numerical resolution, with computational zones as small as 5 pc.
Also with 2D we can compare a large 
number of high resolution calculations with different parameters.
In all calculations discussed below 
the grid is uniform throughout the 
region of interest, usually $z \times R \,=\, 
20 \times 8$ kpc or $40 \times 16$ 
in more extended computations. Beyond the uniform region, 100 zones
of progressively increasing size extend the grid to $\sim 700$ kpc
in both directions. We adopt standard reflecting boundary conditions
at the $R=0$ and $z=0$ edges and outflow boundary conditions at the
outer borders.

Computational difficulties arise when localized, low entropy,
pressure equilibrium perturbations
are imposed in gravitationally bound,
radiating atmospheres like those in galaxy groups
and clusters.
Such perturbations, with lower gas temperature
and higher density, are initially in pressure balance
with the ambient hot cluster gas.
But this equilibrium is immediately broken: 
gravity attracts the denser perturbation
toward the cluster center where the higher ambient pressure 
enhances radiative losses, further amplifying the density 
perturbation.
Idealized coherent perturbations with small or modest amplitude
would undergo radial oscillations at the Brunt-V\"ais\"al\"a
period, having oscillation-averaged cooling
times that are nearly the same as that of the local cluster gas
(e.g. Malagoli, Rosner \& Bodo 1987; Loewenstein 1989).
However, in reality and in numerical computations, perturbations
are never perfectly coherent.
In fixed grid Eulerian computations,
which we consider here, the perturbation moves across
the computational grid.
In general, after each time step. 
as the perturbation moves across the computational grid,  
every grid zone 
has a temperature and density 
that is a mass average of the remaining original gas in that 
zone and gas that has flowed into it.
As a result of this numerical mixing,
the computed rate that gas cools
in the perturbation can either underestimate
or overestimate the radiative cooling that actually occurs.

The mass of gas that cools below some specified low temperature
is computationally underestimated if the internal energy in the
grid-crossing perturbation increases by numerical mixing
with cluster gas faster than it can cool by its own
radiative losses.
As a result of numerical mixing,
the amplitude of a  
moving entropy perturbation can decrease unrealistically fast.
Conversely,
the computed cooling rate can exceed the true cooling rate
if the density contrast in a perturbed zone becomes
sufficiently large relative to nearby cluster gas.
Perturbations that successfully cool on the computational
grid often develop zone to zone density and temperature variations
that are inappropriately large 
for the linearized hydrodynamic 
difference equations (see Koyama \& Inutsuka 2004).
When only a small fraction of the
mass in a dense cooled or cooling perturbation mixes with 
much hotter 
(and therefore less dense) gas in an adjacent zone, 
the numerically averaged temperature is also very low,
causing the hot gas in the 
adjacent zone to cool much too rapidly.
This spurious overcooling by numerical mixing is enhanced by 
the non-linear dependence of radiative
cooling on the gas density and temperature.
Computational experiments show that when the entropy
in a single initial grid zone in a hot atmosphere 
is lowered beyond a
certain threshold, its motion across the grid
stimulates a total mass of cooled gas that can exceed
its initial mass by orders of magnitude. 
As these overdense zones acquire a terminal infall velocity 
relative to the cluster gas, 
pressure gradients in 
nearby, hot gas drive localized radiating inflows  
toward the cooling and infalling zone,
greatly increasing its cooling rate.
Owing to non-linear cooling,
the number of grid zones that cool usually does not
increase; instead the cooled mass increasingly resides in only
a very few zones.
Based on our experience, 
non-linear overcooling zones with ultralow temperatures cannot 
be avoided simply by resorting to smaller computational grids. 

As they fall across the grid moving into 
much hotter ambient gas,
small regions of dense, cooled gas can trigger
large unintended overcooling due to unphysical 
numerical mixing with the hotter gas.
Conversely, in very slowly cooling, low amplitude, 
nearly adiabatic perturbations some degree of undercooling 
can be expected.
Physical truth lies between these two extremes.
In view of these difficulties, we have devised an appropriate 
subgrid scheme to compute radiatively 
cooling gas more accurately.

\subsubsection{Cooling Dropout}

Numerical inaccuracies resulting in overcooling 
can be largely avoided by 
removing strongly cooling gas before it 
stimulates unphysical numerical cooling 
by advecting across the grid. 
To accomplish this, we employ a 
a sub-grid mass dropout procedure 
similar to that described by 
Brighenti \& Mathews (2002).
Our subgrid scheme is similar to early computations of
cooling flows in which gas mass lost by 
star formation was represented 
with a sink term in the continuity equation 
$-q\rho/t_{cool}$ with constant $q$
(e.g. Fabian, Nulsen, \& Canizares 1984; 
White \& Sarazin 1987; 
further references in Mathews \& Brighenti 2003).
Assuming $\Lambda \propto T^{-0.7}$ 
for $10^6 \lta T \lta 10^7$ K,  
the sink term removes gas mass at a rate 
$\rho/t_{cool} \propto \rho^2 / T^{1.7}$. 
We adopt here a modified temperature-dependent 
mass dropout coefficient $q(T)$ that 
increases the sensitivity to cooling gas.
The following mass dropout term is added to
the right hand side of the continuity equation (Eqn. 1):
\begin{equation}
{\left({\partial \rho \over \partial t}\right)_{do}} 
= -q(T){\rho \over t_{cool}}
\end{equation}
(Brighenti \& Mathews 2002).

Since mass removal also results in a loss of thermal energy, 
a corresponding dropout term is also 
needed in the thermal energy equation (Eqn. 3):
\begin{equation}
\left({\partial e \over \partial t}\right)_{do}
= -q(T){e \over t_{cool}}
\end{equation}
For the dimensionless coefficient $q(T)$ we adopt a 
function that increases rapidly with decreasing 
temperature, 
\begin{equation}
q(T) = q_0\exp[-(T/T_c)^2],
\end{equation}
so the dropout term is negligible for $T\gta T_c$.
Accurate mass removal also requires the  
the time step $\Delta t$ be sufficiently small,
allowing the temperature to decline 
slowly over many time steps
even in the most rapidly cooling zones.
Since the exponential term in $q(T)$ 
dominates, the precise value of $q_0$ hardly matters, 
and $q_0 = 2$ is chosen 
to match our previous calculations.
The threshold temperature $T_c$, below which mass dropout
is strongly enhanced, must be small enough to allow accurate
cooling down to temperature $\sim T_c$ 
without significant mass dropout. 
Empirical computations with varying grid resolution 
and $T_c$ converge to the same total mass
of cooled gas provided $T_c \lta 5 \times 10^5$ K.
We note that the sink terms for dropout are
proportional to $t_{cool}^{-1} \propto \Lambda$ 
so that no-dropout occurs in adiabatic flow $\Lambda = 0$.
In addition, high-resolution (small grid zones) provides 
more accuracy in defining the spatial extent of 
strong cooling.

\begin{figure}
\centering
\includegraphics[width=3.in,scale=1.0,angle=0]{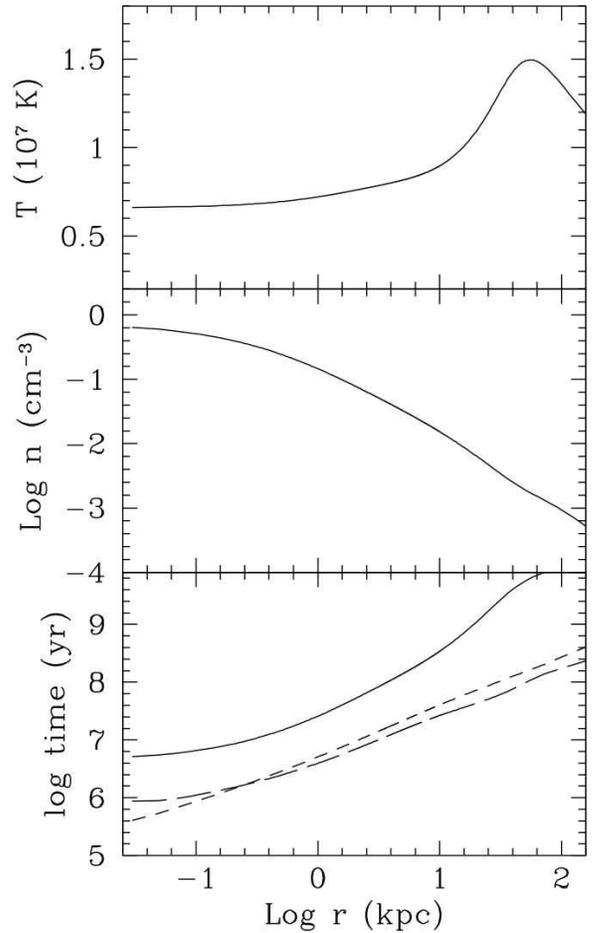}
\caption{
Hot gaseous atmosphere in the NGC 5044 group.
{\it Upper panel:} gas temperature;
{\it Center panel:} gas density;
{\it Bottom panel:} time scales of interest: 
local cooling time (solid); 
local freefall time (short dashed line) 
and the Brunt-V\"ais\"al\"a time (long dashed line).
}
\label{fig1}
\end{figure}

\section{Cooling associated with cavity formation}

We now describe how the evolution of X-ray cavities 
stimulates significant off-center cooling of the 
hot gaseous atmospheres in galaxy groups and clusters. 
For this purpose 
we adopt the well-studied 
atmosphere in the NGC 5044 galaxy group at distance 
31.2 Mpc (Tonry et al. 2001), although our results apply in general to
any central galaxy in groups or clusters.
The atmosphere of this luminous group contains extended 
multifrequency emission believed to result from 
AGN feedback events: filamentary 
optical line emission from warm ($T \sim 10^4$ K) gas 
is observed  
out to $\sim10$ kpc (e.g. David et al. 2011); extended diffuse 
[CII]$\lambda$158$\mu$m emission (Werner et al. 2014);
CO molecular emission within $\sim$4 kpc, 
much of which is contained in 
about two dozen small clouds (David et al. 2014), 
and 70$\mu$m dust emission out to $7-10$kpc where the 
sputtering time is only $\sim 10^7$yrs
(Temi, Brighenti \& Mathews 2007a,b). 
This dust has probably been transported outward 
from the core by AGN events (Mathews et al. 2013). 
The central AGN in NGC 5044 is visible as an X-ray 
point source. 

\begin{figure*}
\centering
\includegraphics[width=5.5in,scale=1.0,angle=-90]{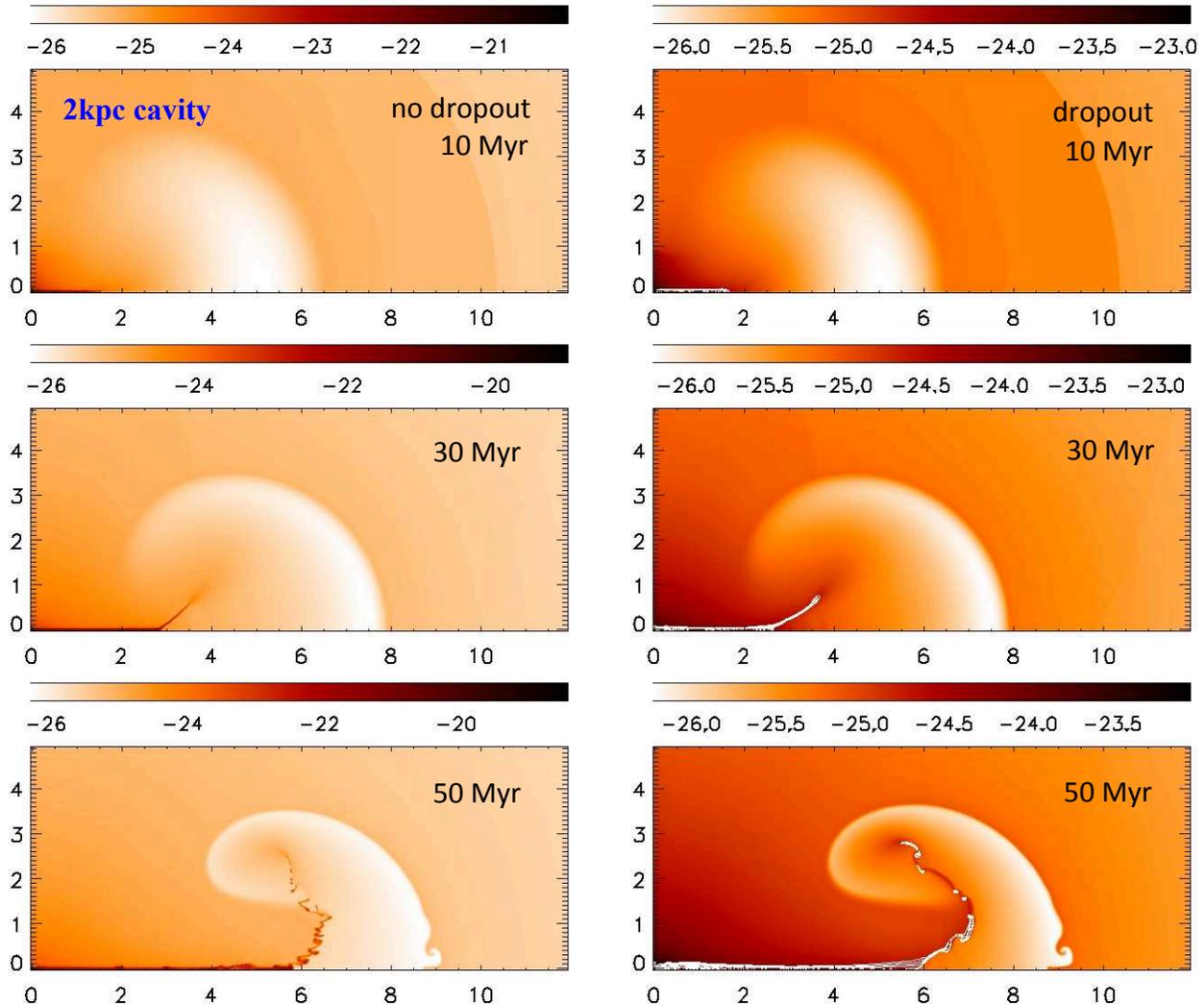}
\caption{
Gas density maps at various times for the 2 kpc cavity calculated
without dropout (left panels) and with dropout (right panels). In the
left panel, dark regions represents 
cooled gas at $T=10^4$ K. In the right
panels, white grid zones indicate locations where the gas is
currently dropping out of the flow (cooling to 
temperatures below $T_c = 5 \times 10^5$K).  
Axes in each panel in this and subsequent figures are labeled with $z$ 
(horizontal) and $R$ (vertical) coordinates in kpc.
}
\label{fig1}
\end{figure*}

The upper panel in Figure 1 shows 
our approximate fit to the temperature 
profiles from X-ray observations of NGC 5044 
(David et al. 1994; Buote et al. 2003,2004; David et
al. 2009) and the density (central panel) is determined from  
X-ray observations by 
assuming hydrostatic equilibrium. 
The adopted gravitational potential is the combination of a dark
matter NFW halo (Navarro et al. 1996) (mass $M_{\rm dm}=4\times
10^{13}$ M$_\odot$ and concentration $c=8.8$) and a de Vaucouleurs
galaxy of mass $M_*=3.4\times 10^{11}$ M$_\odot$ and
effective radius $r_{\rm e}=10$ kpc.
The bottom panel of Figure 1 shows several time scales of
interest: the cooling time at constant pressure
$t_{cool} = 5 m_p kT/2\mu \rho \Lambda$,
the local freefall time
$t_{ff} \approx (|g|/2r))^{-1/2}$
and the Brunt-V\"ais\"al\"a time
$t_{bv} \approx [|g(d\ln \rho /dr)|]^{-1/2}$.
Hydrostatic equilibrium is assumed to hold in the
initial spherical hot atmosphere,
$g = (1/\rho) dP/dr$, where $r$ is the radial coordinate.

\begin{figure*}
\centering
\includegraphics[width=6.5in,scale=1.0,angle=-90]{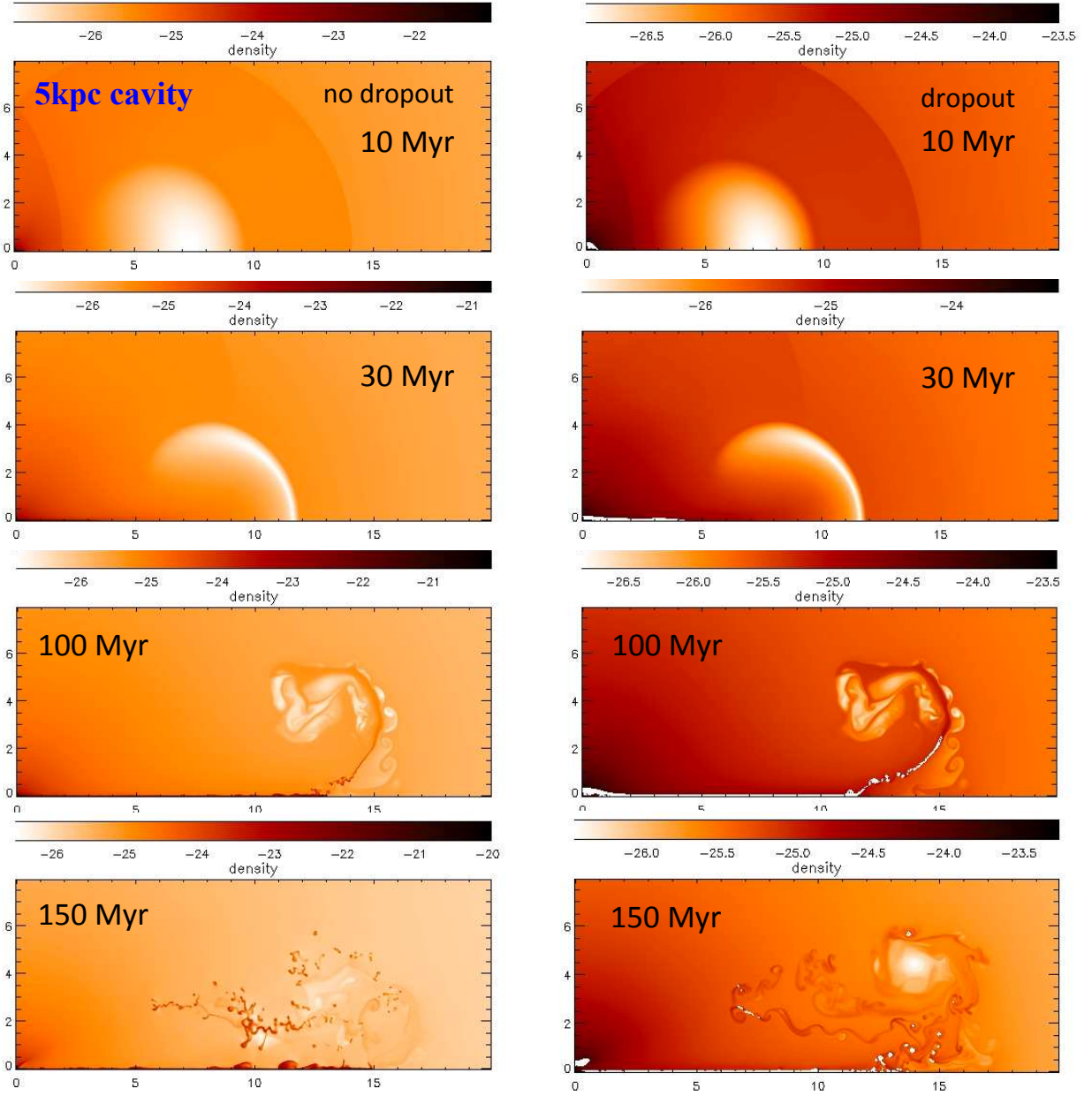}
\caption{
Same as Figure 2, but for the 5 kpc cavity.
}
\label{fig1}
\end{figure*}

All hydrodynamic calculations begin with a static 
spherical atmosphere having 
density and temperature profiles as in Figure 1. 
Immediately as the calculation begins,
a cooling flow develops in the hot atmosphere 
and a cavity is created by heating 
gas in a small spherical region along the $z$-axis, 
generating an expanding cavity that drives a modest shock
wave into the ambient gas. 
The heating continues for $t_{heat} = 1$ Myr
and provides a total internal energy 
$E = 1.17 \times 10^{57}$ ergs to the atmosphere, 
corresponding to a power of $3.7 \times 10^{43}$ erg s$^{-1}$. 
This short cavity formation time is consistent with the 
absence of observable jets 
in most galaxy group cavities. 
In reality cavities are thought to be inflated by 
relativistic and/or ultra-hot gas transported 
from the central AGN by jets.
Nevertheless, our conclusions about cooling during 
cavity evolution are quite 
general and are quite independent of the particular means 
of cavity inflation.
In the discussion that follows 
cavities are defined by their
fixed heating points in the NGC 5044 atmosphere, e.g. 
``2 kpc'' or ``5 kpc'' cavities.

\begin{figure*}
\vskip-0.5in                                                                               
\centering
\includegraphics[width=6.5in,scale=1.0,angle=0]{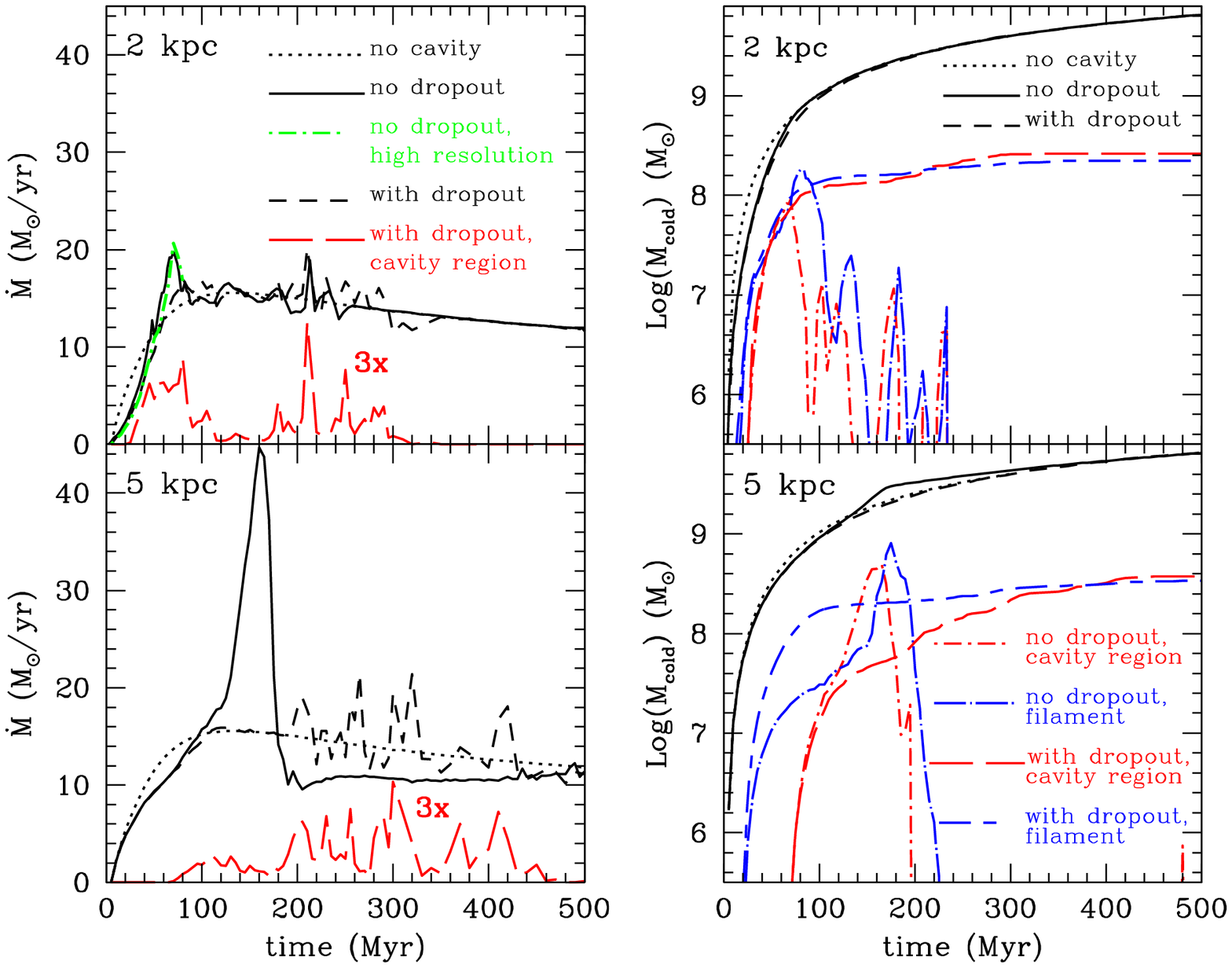}
\vskip-0.9in
\caption{
{\it Left panel:} Time variation of cooling rates 
${\dot M}$ for the 2 kpc cavity (upper
panel) and 5 kpc cavity (lower panel). 
Dotted lines show the global cooling with no cavity.
Solid lines represent models 
without dropout (i.e. gas cools to $10^4$ K and stays on the
grid). Dashed lines show models with dropout (in which
the gas that cools to $T\sim T_{\rm crit} = 5\cdot 10^5$ K is
smoothly removed from the grid).  
Long dashed lines at the bottom of each panel 
plot three times the rate that 
gas drops out in the cavity region ($z > 1$ kpc $R > 0.3$ kpc).
The green dash-dotted line (extending only to 80 Myr)
shows a repeat of the no-dropout 
2 kpc evolution at ultra-high grid resolution (5 pc), 
$\Delta z = \Delta R = 5$ pc.
{\it Right panel:} Time variation of the mass of cold gas 
(either cooled to $T = 10^4$ K or dropped out). 
Dotted black lines near the top show the total mass of gas 
that cools at the center of the atmosphere when no cavity forms.
Solid and dashed black lines near the top 
represent the total mass of cold gas without and with dropout 
respectively. 
For dropout flows: 
long dashed and long-short dashed lines are cumulative dropped 
out mass for the ``cavity region''
($z > 1$ kpc, $R>0.3$ kpc) 
and the ``filament region'' ($z > 1$ kpc, $R<0.3$ kpc).
For no-dropout flows: long-dash-dotted and short-dash-dotted lines 
are the total cold mass in the filament 
and cavity regions respectively.
Blue lines denote cold gas in
the ``filament region'' ($z > 1$ kpc, $R<0.3$ kpc); 
red refers to 
cold gas in the ``cavity region'' ($z > 1$ kpc, $R>0.3$ kpc).
}
\label{fig1}
\end{figure*}

\subsection{Evolution of the 2 kpc and 5 kpc cavities}

Figures 2 and 3 show the gas density evolution of 2 kpc and 5 kpc 
cavities at selected times. 
Although the total thermal energy that formed both cavities 
is the same, their maximum transverse cavity radius increases from 
3.4 to 4 kpc as the heating point moves from 2 to 
lower density gas at 5 kpc 
from the cluster center. 
Solutions with and without mass dropout are shown respectively in 
right and left panels of Figures 2 and 3. 
If there is no mass dropout ($q(T) = 0$), 
gas cools to $T = 10^4$ K then remains on the 
grid, supposedly ionized by UV from old asymptotic branch giant stars, 
i.e. $\Lambda(T) = 0$ for $T < 10^4$ K.
This type of arrested cooling endstate
is plausible since as the gas cools, small magnetic
fields in the initial gas increasing as $B \propto \rho^{2/3}$  
can ultimately provide enough magnetic pressure to balance the 
pressure of local hot gas
(Fabian et al.2008).
Observations of the density-sensitive [SII]$\lambda$6716/6731 flux
ratio in extended warm ($T \sim 10^4$ K) gas 
around M87 (Werner et al. 2013) 
indicate that the density increase in cooling gas 
may be magnetically arrested. 
When mass dropout at low temperatures is present 
and $T_c = 5 \times 10^5$K, 
as in the right hand panels in Figures 2 and 3, 
we show with white contours grid zones in which gas 
is currently dropping out. 
In dropout solutions 
cooled gas is removed from the grid and its subsequent 
dynamical motion is not followed. 
All cavity evolutions in Figures 2 and 3 
were made at high spatial resolution 
with uniform grid zones of size $\Delta z = \Delta R = 20$ pc, 
i.e. each square kpc in the Figures contain $50 \times 50 = 2500$ zones.

While the general appearance of cavities with and without
mass dropout in Figures 2 and 3 are nearly the same, 
a detailed analysis of the cooling rate and the total accumulated mass
of cooled gas shown in Figure 4 reveals significant differences.
Several cooling rates are plotted in the left panels in Figure 4, 
describing the total cooling rate of cluster gas in the absence of 
cavity formation (dotted lines) and total cooling with
(dashed lines) or without (solid lines) the dropout terms.
(The long-dashed lines plotted near the bottom in the left
panels are cooling rates in the ``cavity region''
and will be discussed further below.)

The single cavities we consider do not provide enough 
distributed long-term energy 
to stop radiative cooling near the center 
of the flow; to accomplish this would require more cavities 
with various locations and energies (Mathews 2009) 
which would mask and complicate the single cavity 
flows we describe here. 
Consequently, as seen in Figure 4, 
much of the cooling and cooled gas 
accumulates right at the center. 
During the initial 70-100 Myrs, as the 
cavities form and the atmosphere converts to a cooling flow, 
the cavities are seen to slightly reduce the total cooling 
rate.
This reduction is apparently 
due to heating by the cavity shock 
and the large vortical flow generated 
around the buoyant cavity. 
As the outward moving cavity provides space to be filled 
in its wake,
cluster gas flows down around the outer cavity boundary 
then circulates up from beneath the cavity, 
joined by low entropy gas closer to the cluster center 
that would have otherwise flowed to the center and cooled.
The total cooling rate ${\dot M}$ is reduced by this outward 
motion of low entropy gas in the cavity wake.

\begin{figure}
\vskip.15in
\hskip-0.3in
\includegraphics[width=3.5in,scale=0.8,angle=0]{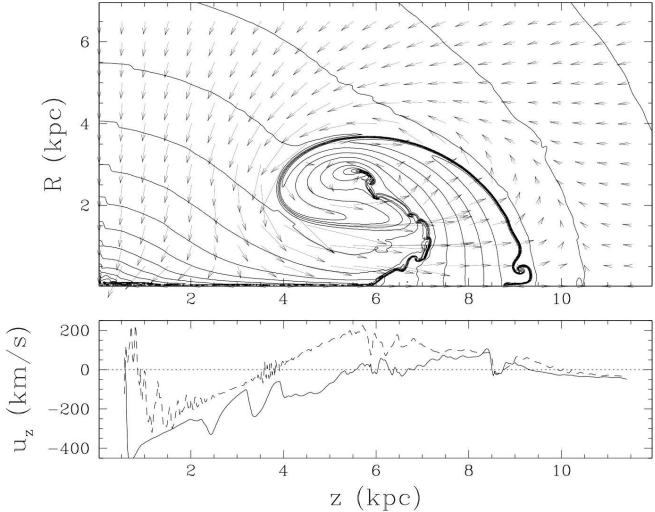}
\caption{
{\it Upper Panel:} Velocity field for dropout flow for 
2 kpc cloud at 50 Myr. Arrows 1 kpc in length 
correspond to 200 km s$^{-1}$. 
Contours are for gas density.
{\it Lower Panel:} Radial velocity along $z$-axis for 2 kpc cavity
at time 50 Myr.
Solid and dashed lines are for no-dropout and dropout flows respectively.
}
\label{fig2}
\end{figure}

\subsection{The cavity vortex flow}

A snapshot of the velocity field at 50 Myrs in the 
buoyant 2 kpc cavity for the dropout 
solution is shown in the upper panel of Figure 5.
The large vortex flows in a slanted fashion 
toward the symmetry $z$-axis 
and broadly up into the bottom of the cavity.
This broader upflow provides 
a sustained compression in which much of 
the radiation and entropy losses occur  
during the early cavity evolution.
Solutions without dropout are seen to 
undergo periods of intense cooling 
near 70 and 160 Myrs respectively 
for the 2 kpc and 5 kpc cavities (solid lines in Fig. 4, left
panels). For both cavity models much cooling occurs 
in the ``cavity region'' away from the center or $z$-axis filament, 
as seen in the left panels of Figure 4.
These peaks in the total cooling rate 
are due to numerical overcooling triggered by 
numerous cold clouds,
clearly visible in the 
lowest left panel of Figure 3 for the 5 kpc cavity model.
However, these peak cooling rates are entirely absent
when dropout is included (dashed lines in Fig. 4). 
The largest of these spurious cooling peaks in Figure 4, shown 
in the 5 kpc cavity at time 150 Myr (bottom left in Figure 3),
occurs at a time when the cavity has largely disappeared 
and the off-axis no-dropout flow is dominated by 
many small and dense overcooling regions 
falling through the hot gas (and across the computational grid).
So much gas overcools during the spurious 5 kpc cavity 
cooling peak in Figure 4 
that the subsequent global cooling rate does not recover 
until about time 500 Myrs.

The radial velocity $u_z(z)$ 
along the $z$-axis filament for the 2 kpc 
cloud at 50 Myr is plotted in 
the lower panel of Figure 5 for the dropout (dashed line) and 
no-dropout (solid line) solutions.
The no-dropout solution is more irregular because 
small, dense clouds with temperature $T = 10^4$ K 
continue to collide into the filament.
These dense regions, usually only several 
grid zones in size, do not free fall in a perfectly radial 
direction toward the center of the flow, 
but are swept toward the filament along the $z$-axis 
by the hot gas vortex flowing beneath the cavity. 
The filament in the no-dropout solution, 
also at $T = 10^4$ K, 
is denser and flows toward the center more rapidly 
than the filament in the dropout solution 
where hot gas is still cooling at temperatures $T \gta 10^5$ K.
Note that the filament velocity $u_z(z)$ passes 
through zero.
The zero-velocity point typically 
moves out with time toward larger $z$ 
(Mathews \& Brighenti 2008).

\subsection{Off-center cooling beneath cavities}

The uppermost solid (no-dropout) and dashed (dropout) lines in the
two right hand panels in Figure 4 show the total cumulative mass 
that either cooled or dropped out during the first 500 Myrs.  
The total cooled mass, $6 \times 10^9$ $M_{\odot}$,
is nearly unaffected by dropout, initial 
location (2 or 5 kpc) of the cavity, 
or whether the cavity is present or not.
Most of this global cooling occurs at 
the origin \footnote{In the no-dropout case the
cooled gas accumulates at the center, where it stays indefinitely.}
and could be reduced or eliminated by multi-cavity feedback events.
Our interest here is the mass of non-central  
gas that cools along the 
$z$-axis filament and, especially, 
away from both the filament and the center.

\begin{figure}
\centering
\includegraphics[width=3.2in,scale=0.8,angle=0]{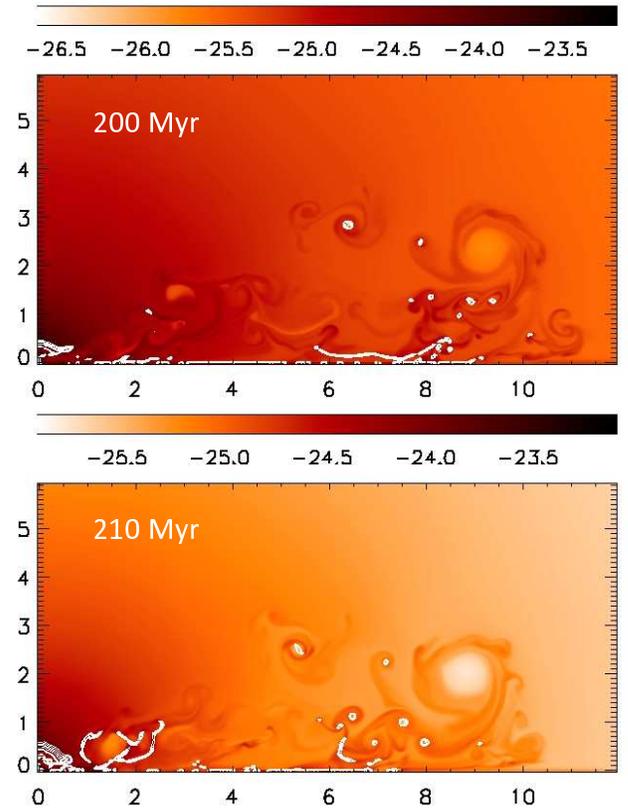}
\caption{
Density images for 2 kpc cavity 
dropout flow at 200 and 210 Myrs.
}
\label{fig2}
\end{figure}

To explore off-center cooling, 
we consider non-central cooling in 
two regions: the ``filament region'' along the $z$-axis 
($z > 1$ kpc $R < 0.3$ kpc) and the ``cavity region'' 
($z > 1$ kpc $R > 0.3$ kpc). 
Cooling in the radial filament and cavity regions 
are shown respectively
with blue and red lines in Figure 4 or in terms of 
line types as explained in the figure caption.

\begin{figure*}
\centering
\includegraphics[width=6.5in,scale=0.8,angle=0]{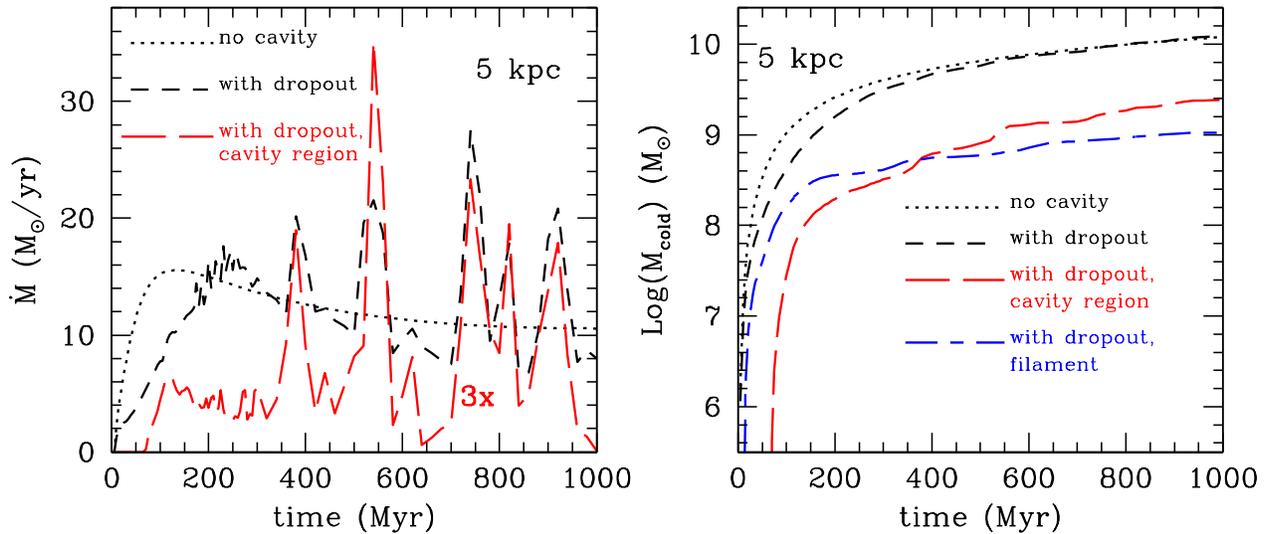}
\vskip.3in
\caption{
Cooling rate and total cumulative cooling 
for 5 kpc cavity with 10 times higher energy 
$1.17 \times 10^{58}$ ergs.
Line types and colors are the same as in Figure 4.
}
\label{fig2}
\end{figure*}

The episodic cooling pattern during the no-dropout cavity 
evolution (right panel of Fig. 4) clearly indicates that some of the 
gas cools first in the cavity region 
then flows into the filament region, 
although much cooling occurs in gas flowing 
directly into the filament region.
After gas enters the filament, it eventually flows with nearly freefall
or terminal velocity into the core of the cluster-centered galaxy. 
This filament persists long after the cavity has 
become difficult to observe, $t \gta 70$ or 160 Myrs respectively for 
the 2 kpc and 5 kpc cavities. 
(The visibility of coherent cavities in Figures 2 and 3 may 
be degraded somewhat by residual numerical mixing.)
Such radial filamentary features 
have been observed in warm gas emission in many groups and clusters.
The perfectly focused, narrow width of the filaments we compute 
is due to the 
exactly static cluster gas prior to forming the 
axisymmetric cavity.
In reality, the cluster gas is expected to have random 
subsonic motions due to feedback from previous generations 
of cavities, and this initial turbulence would
deflect gas as it approaches the $z$-axis,
broadening the filaments.
In no-dropout calculations gas that cools episodically  
in the cavity region later flows into the filament region then 
into the origin, accounting for the time lag between cooling 
peaks in cavity and filament regions in Figure 4.

\begin{figure}
\centering
\includegraphics[width=2.8in,scale=0.8,angle=0]{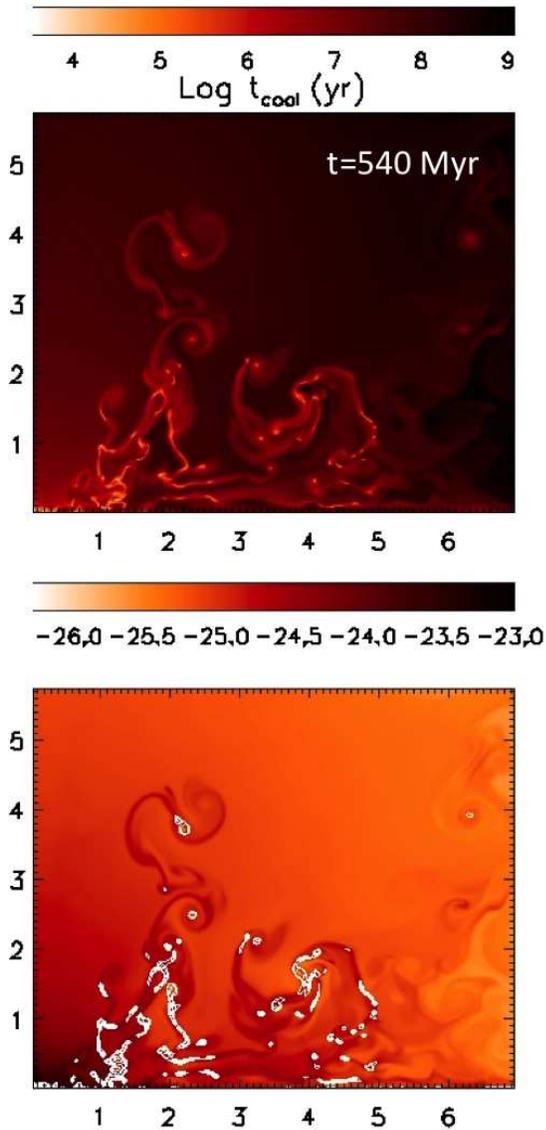}
\vskip.3in
\caption{
Cooling time $t_{cool}$ ({\it top}) and 
gas density with currently dropping-out zones in white 
({\it bottom}) 
near the center of the hot atmosphere  
for the high energy 5 kpc cavity 
at time t = 540 Myr.
}
\label{fig2}
\end{figure}

Red long dashed and blue long-short dashed 
lines in the right panels of Figure 4 
show the total cumulative mass of gas that cooled or dropped out 
respectively in the cavity and filament regions.
Slope changes (as in the red dashed line 
for the 2 kpc cavity at $t \approx 210$ Myrs) 
indicate episodes of more intense cooling. 
Since we do not follow the dynamics of gas after it 
drops out, it is not possible to know where this gas is currently 
located, but as in the no-dropout case, most of it ultimately 
find its way to the origin or form stars. 

The total mass of gas that cools off-center varies with time,
but can be large. 
Dropout cavity calculations do not suffer from overcooling 
and the amount, time and location of gas that drops out 
is recorded during each dropout evolution.
The cooling (or dropping out) rate ${\dot M}_{cav}(t)$ of gas 
in the cavity region 
($z > 1$ kpc $R > 0.3$ kpc) and the total mass of dropped out gas
are plotted with dashed lines in the left panels of Figure 4,
where for visibility ${\dot M}_{cav}(t)$ is multiplied by 3.
The 2 kpc cavity experienced sustained compression 
in the vortical upflow lasting about $\delta t = 25$ Myr 
from $\sim$40 to $\sim$65 Myr.
During this time, gas beneath the cavity is compressed sufficiently
long to radiate away a substantial fraction of its entropy 
or to cool completely. 
Ongoing cooling of this gas is visible 
as white cooling regions dropping out in the density image 
for the 2 kpc cavity at time 50 Myrs in Figure 2.
The total mass that cools during this cavity compression 
episode is the product 
of the dropout rate 
${\dot M}_{cav} \approx 2$ $M_{\odot}$ yr$^{-1}$ 
and the duration, 
${\dot M}_{cav} \delta t \approx 8 \times 10^7$ $M_{\odot}$.

We note here that in real systems X-ray cavities are continuosly
created by the central AGN. While the effect of recurring cavities on the
off-center cooling will be considered in a future work,  we point out
that preliminary calculations validate the fundamental result
described above. Intermittent cavities persistently generate regions
of converging flow and non-linear density perturbations which
originate spatially extended cold clouds (see also Brighenti \&
Mathews 2002).

\subsection{Late time cooling}

In addition to cooling during the initial 
cavity compression, Figure 4 indicates that 
the mass dropout rate in the cavity region 
${\dot M}_{cav}(t)$ continues to 
peak in an irregular fashion at much later times.
In fact, 
the lower left panel of Figure 4 shows that the 5 kpc cavity 
cools considerably more mass at times $t \gta 200$ Myr
after the cavity has largely disappeared 
than it does during the initial cavity compression
at time $t \sim 100$ Myr. 

The 2 kpc cavity also exhibits a continued high rate of 
intermittent cooling until 300 Myrs.
Consider for example 
the large late time cavity region mass dropout 
in the 2 kpc cavity event at $\sim 210$ Myr.
This dropout event 
at $(R,z) \approx (0.5,1.5)$ kpc 
lasts  $\delta t \approx 12.5$ Myr 
with mean ${\dot M}_{cav} \approx 2$ $M_{\odot}$ yr$^{-1}$, 
corresponding to a total cooled mass 
$\sim 2.5 \times 10^7$ $M_{\odot}$. 
Figure 6 shows the spatial distribution of this cooling gas 
at 200 Myrs just before this cooling event and 
at 210 Myrs when it peaked.
The peak in the
cavity region ($z > 1$ kpc $R > 0.3$ kpc) at 210 Myrs 
is apparently due to the irregular white features 
seen in Figure 6 at $1.5 \lta z \lta 2.5$ kpc.
Evidently, cluster gas that experienced a relatively small
amount of entropy loss during compressions in 
the early cavity evolution 
eventually flows slowly inward and cools  
away from the cluster center.
In pressure equilibrium gas with lower entropy 
has a higher density and sinks in the atmosphere.
The complex cooling pattern that appears 
as curved filaments when viewed in the 
axisymmetric density slice in Figure 6  
is actually toroidal. 
Nevertheless, it still has a filamentary 
appearance when viewed in projection.
Many such curved, non-radial warm gas 
filaments have been observed in optical images 
in thermal line emission 
(e.g. Conselice et al. 2001; McDonald et al. 2012).

When observing the NGC 5044 group with the ALMA array, 
David et al. (2014) discovered about 2 dozen
gravitationally unbound CO(2-1) molecular clouds 
having individual masses of 
$10^5$ to $10^7$ $M_{\odot}$ with a combined total mass 
of $5 \times 10^7$ $M_{\odot}$. 
Additional, more diffuse CO(2-1) emission 
within the central 1.6 kpc 
may have been resolved out by ALMA, as indicated 
by single dish IRAM 30m CO(2-1) observations that detect a total 
molecular mass in NGC 5044 of $\sim 2 \times 10^8$ $M_{\odot}$.
These molecular masses are estimated by assuming that the Galactic 
CO to H$_2$ conversion is appropriate. 
The array of cooling gas at $(R,z) \approx (0.5,1.5)$ kpc
seen in the lower panel of Figure 6 
has enough mass ($\sim 2.5 \times 10^7$ $M_{\odot}$) 
to create 25 molecular clouds each of mass $10^6$ $M_{\odot}$ 
within 2 kpc from the center, 
attributes similar to the clouds 
observed in NGC 5044 by David et al. (2014).

If molecules form in the late time cooling gas in our calculations, 
the resulting clouds would be unbound, dust-free and 
rather randomly distributed in the central few kpcs. 
In these respects they resemble the CO clouds discovered
in the NGC 5044 atmosphere by David et al. (2014).

Long dashed lines in the right panels of Figure 4 indicate that 
the total cold, dropped out 
mass in the cavity region 
due to the evolution of the 2 and 5 kpc cavities over
500 Myrs are 
$2.6 \times 10^8$ $M_{\odot}$ and $3.5 \times 10^8$ $M_{\odot}$
respectively. 
More gas cools in the cavity region 
when the same cavity energy is injected 
further out in the atmosphere, creating larger cavities.
The total mass cooled over time by the 2 or 5 kpc cavities 
is comparable with the total molecular mass 
observed in NGC 5044 ($\sim 2 \times 10^8$ $M_{\odot}$). 
Because of intermittent cooling,
not all of the computed dropped out gas would be 
visible at any one time, so multiple or more energetic cavities 
would be required
to continuously supply the CO(2-1) mass observed.

\subsection{A more energetic cavity at 5 kpc}

Figure 7 shows the cooling rate and total cooled gas 
identical to the 5 kpc cavity discussed earlier 
but now having 10 times more energy, $1.17 \times 10^{58}$ ergs.
Cooling in the cavity region continues to 
at least 1 Gyr when our calculation terminated.
By 1 Gyr $2.5 \times 10^9$ $M_{\odot}$ of 
gas cools in the cavity region at a remarkably high
average rate ${\dot M}_{cav} = 2.5 M_{\odot}$ yr$^{-1}$.
After 300 Myrs, when the cavity has largely disappeared,
the cooling rate is dominated by 5 major events each lasting 
$\sim50$ Myrs when the cooling rate 
${\dot M}_{cav}$ exceeds $4 M_{\odot}$ yr$^{-1}$.
During the largest cooling event at time 540 Myrs 
lasting $\delta t \approx 40$ Myr, a total of 
$2 \times 10^8$ $M_{\odot}$ cools in the cavity region.
As seen in Figure 8, most of this gas cools within 
5 kpc of the center, typical of late time cooling.
The fate of the gas that cools in this dropout 
calculation is unclear, but it 
may remain as warm or molecular gas for a long time 
after it cools by dropout so it may be possible 
to observe this cold gas long after each peak in the 
cooling rate in Figure 7.

\subsection{Velocity distribution of the cold gas}

It is interesting to investigate the velocity distribution of the cold
gas generated by the ''cavity-cooling'' process. In the following we
focus on the cold clouds in the cavity region. We ignore the cold
filament on the $z-$axis because the imposed symmetry of our 2D
cylindrical calculations might result in enhanced on-axis cooling.
Moreover, we neglect the effect of the relativistic jet, from which the
cavity likely originates, that can modify the velocity structure of
the gas on the axis.

Figure 9 shows the distribution of the $z-$component of the cold gas
velocity for several no-dropout models, in the cavity region. 
Generally, both inflowing and outflowing cold
clouds are present, with no clear trend. 
Cold outflows form from outward moving hot gas and
are not clouds dragged from the center by the cavity-induced motion
(see also Costa et al. 2014). Most cold gas has $v_z$ in the range
[-200, +200] km s$^{-1}$, although high velocity outflows with $v_z
\sim 400$ km s$^{-1}$ are sometime present (Figure 9) We do not find an obvious
correlation between maximum outflow velocity and the power of the
cavity. The model with $E_{cav} = 1.17\times 10^{58}$ ergs (Section
3.5) generates cold outflows with velocity comparable to the the model
wih $E_{cav} = 1.17\times 10^{57}$ ergs  and cavity at 2 kpc. 
At late times the cold gas, which is always bound to the system, tends to
fall to the center and the velocity distribution becomes slightly
skewed toward negative values.

Recent ALMA observations of massive central galaxies (Russell et al. 2014,
McNamara et al. 2014), show molecular gas with velocity up to $\sim
500$ km s$^{-1}$. If these velocities are representative of outflowing
material, it is likely that multiple cavities or AGN
outflows are necessary to explain these properties.
Valentini \& Brighenti (2015) discuss in more detail the velocities
and masses of cooled gas stimulated by many recurrent outflows and
cavities in the same hot gas atmosphere.

\begin{figure}
\includegraphics[width=3.2in,scale=0.9,angle=0]{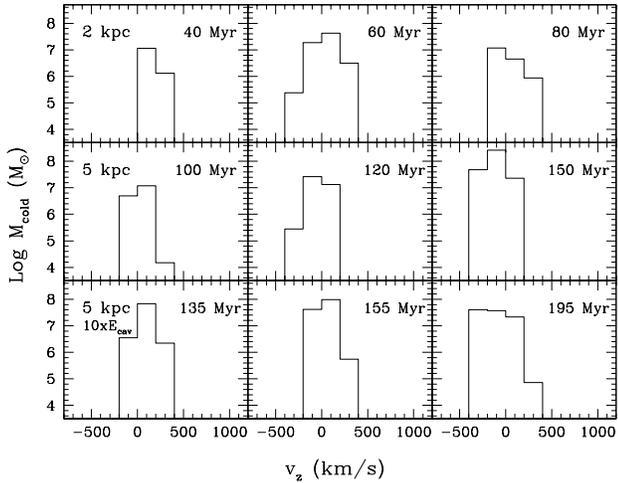}
\caption{
Distribution of the cold gas $z-$component velocity for three
no-dropout models in the cavity region
at several times. Positive velocities indicate outflowing gas. In the
top row is shown the model with the cavity at $z=2$ kpc
($E_{cav}=1.17\times 10^{57}$ ergs). The middle row illustrates the
model with the cavity at $z=5$ kpc ($E_{cav}=1.17\times 10^{57}$
ergs). The bottom row shows the cavity model with $E_{cav}=1.17\times
10^{58}$ ergs, located at $z=5$ kpc.
}
\label{fig9}
\end{figure}

\begin{figure}
\centering
\includegraphics[width=3.2in,scale=0.7,angle=0]{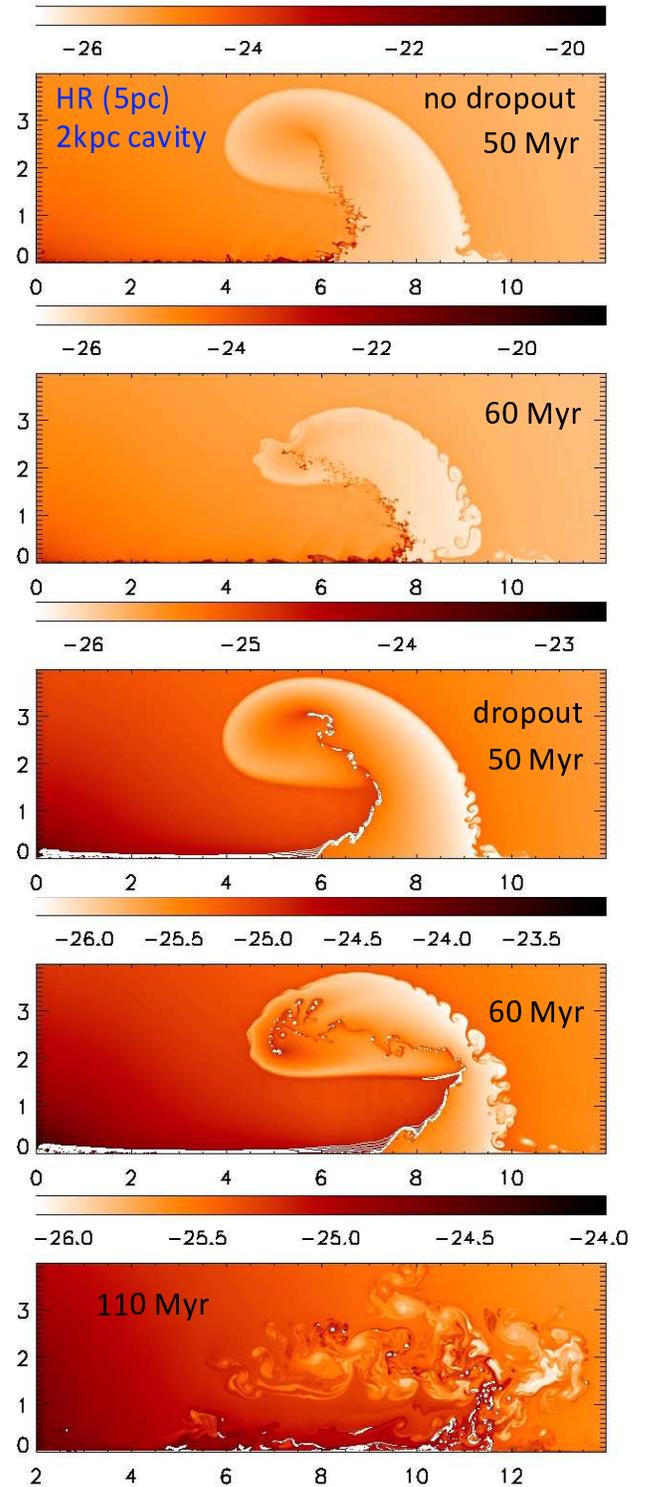}
\vskip.3in
\caption{
Ultra-high resolution ($\Delta z = \Delta R = 5$ pc) 
density images of the 2 kpc cavity 
at 50 and 60 Myrs with and without dropout.
}
\label{fig2}
\end{figure}

\begin{figure}
\centering
\includegraphics[width=3.4in,scale=0.8,angle=0]{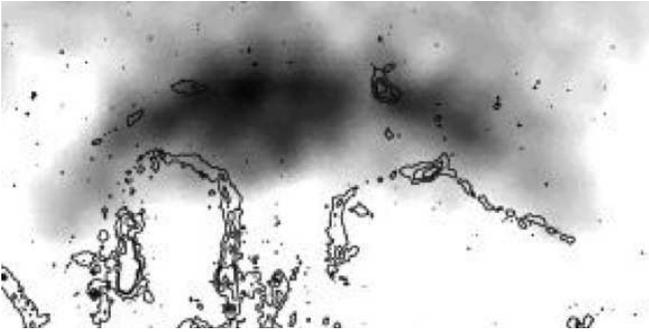}
\vskip.1in
\caption{
The buoyant and flattened 
northwestern cavity in the Perseus cluster 
(shown darkened) is 
rising from the cluster center about 2 cavity 
diameters vertically below the figure 
(Fabian et al. 2003).
Contours show  
H$\alpha$ emission below the cavity that resemble the cooling 
regions computed here.
}
\vskip0.1in
\label{fig2}
\end{figure}

\subsection{Ultra-high resolution}

To demonstrate computational convergence, 
some of the solutions shown in Figures 2, 3 and 6 
with grid size $\Delta z = \Delta R = 20$ pc have been repeated 
at even higher numerical resolution,
$\Delta z = \Delta R = 5$ pc.
Figure 10 shows five ultra-high resolution density images of the 
2 kpc cavity evolution with and without 
cooling dropout.
At 50 Myrs the first and third panels in Figure 10 closely  
resemble the lowest two panels in Figure 2 
and the density snapshots are similar for 
all four images at 50 Myrs, dropout or no-dropout. 
But shortly thereafter at time 60 Myrs,  
the morphology seen in the ultra-high resolution buoyant 2 kpc 
cavity becomes sensitive to the dropout assumption. 
(A similar drastic change occurs during 50-60 Myr at 
the previous 20 pc grid size.)
Between 50 and 60 Myrs 
the no-dropout 2 kpc cavity shrinks in the $R$-direction, 
possibly due to hot gas cooling onto many 
small, dense overcooling regions seen across the base  
of the rising cavity.
During this same time interval, the $R$ dimension of the dropout 
flow remains unchanged, but the vortical upflow pinches 
the buoyant 5 kpc cavity in the $z$-direction.
Soon afterward, the pinched-off part of the cavity will
become a torus.
This pinching is less pronounced in the no-dropout 
solution possibly because of the combined inertia of dense cold 
gas remaining at the bottom of the cavity at $t = 60$ Myr.
We draw attention to the Rayleigh-Taylor and 
Kelvin-Helmholtz features seen in Figure 10 along 
the front and side of the cavity at time 60 Myr. 
The lowest dropout panel in Figure 10 shows the
disrupted 2 kpc cavity at 110 Myrs.
Finally, the green long-short dashed line in Figure 4 
shows the no-dropout cooling rate for the 2 kpc cavity 
at ultra-high 5 pc resolution which 
faithfully follows the unphysical overcooling peak 
of the $\Delta z = \Delta R = 20$ pc solution (solid line).
This agreement not only verifies computational convergence, 
but also shows that the unphysical overcooling in this peak cannot 
simply be removed with higher computational resolution 
or by employing adaptive mesh refinement (AMR).

\subsection{Similar cooling filaments in the Perseus cluster}

In support of the early time cavity 
cooling patterns illustrated in Figures 2, 3 and 7, 
we show in Figure 11 a remarkably similar 
H$\alpha$ filament observed in the 
northwestern cavity in the Perseus cluster 
(Fabian et al. 2003).
The flattened mushroom shape of this cavity indicates 
that is buoyantly rising in the cluster gas. 
We propose that these observed filaments are 
edge-brightened toroidal features that qualitatively
resemble the toroidal 
cooled gas filaments we calculate here. 
These curved filaments in Perseus do not indicate that 
the much colder and denser H$\alpha$ emitting gas 
is ``being dragged'' 
outward in the atmosphere by the rising cavity, 
as many authors have suggested, nor do 
the $\Upsilon$-shaped filaments represent stream lines 
as the gas flows underneath the cavity.
Instead, when interpreted with our calculations,  
the curved filaments in Figure 11 
define a transient cooling 
locus where the vortical upwelling 
of cluster gas presses against the bottom of the rising cavity. 
However, the lower approximately radial 
segments of the Perseus filaments remain parallel and 
open along the symmetry
axis pointing toward the cluster center. 
This feature differs from our computed $\Upsilon$-shaped filaments 
in which the vertical elements of the filament are 
coincident along the $z$-axis.
The separation of the parallel vertical segments of 
the H$\alpha$ filament 
observed in Perseus may be due to the influence of 
the jet that created the cavity
which we do not consider.

\section{Summary and Conclusions}

Confidence in computations of Eulerian cavity hydrodynamics
with radiative losses in group/cluster atmospheres
is only possible when several numerical pitfalls are avoided.
It is essential to avoid overcooling when small,
high density cooled regions move across the computational grid,
seeding additional unphysical
overcooling which can greatly exceed the mass of
the original cooled regions.
We show that numerical overcooling
can be avoided by invoking subgrid cooling dropout
in which gas is removed from the computational grid
before it cools below some very low temperature.
Spurious overcooling is also limited computations with 
high numerical resolution, 
Time steps must be shortened
to smoothly follow the deceasing temperature
at every cooling site in the flow.
The volume occupied by gas that cools to low temperatures
is insignificant and its removal does not upset the overall
dynamics of the remaining hot gas.
The same numerical thermal mixing responsible for overcooling can
lead to undercooling in localized regions
having only slightly smaller entropy than the ambient gas.
In this case the entropy in the perturbation
may rise faster by numerical mixing than it 
decreases by radiative losses, leading to an underestimate
of the cooling rate.
Numerical undercooling is more difficult to detect and avoid
than overcooling,
but when radiative cooling occurs in computations with dropout,
this provides a physically useful lower limit 
on the total mass of gas that actually cools.

Detailed high resolution computations of the rise and demise 
of buoyant cavities demonstrate beyond doubt that 
cavities visible in X-ray group/cluster atmospheres 
stimulate inhomogeneous, spatially extended, off-center 
cooling of significant masses of hot gas. 
Cooling occurs in an ordered fashion during the early  
evolution of cavities 
and later in a more disordered fashion 
long after the initially spatially coherent cavities have 
irreversibly fragmented and disrupted. 
Cavity-induced cooling is expected to be 
universal in hot gaseous  
atmospheres in galaxy groups and clusters,
but it may not always be observable 
due to variations in the cooling rate.
To illustrate cooling associated with buoyant cavities, 
we compute cavity hydrodynamics in the well-studied 
hot atmosphere of the NGC 5044 group which is known 
to have three spatially extended multi-phase components: 
optical line emission from warm gas, 
far infrared dust emission, 
and submillimeter emission from molecular gas.
Apparently, most of 
the observed low temperature gas currently in NGC 5044 is molecular.

To investigate off-center cooling, 
we consider the inflation of single axisymmetric cavities 
in an idealized, perfectly static 
hot gas atmosphere resembling NGC 5044, 
The earliest off-center cooling appears 
beneath the cavity along the symmetry axis.
Shortly afterward, cooling appears across the bottom 
of the buoyant cavity as it flattens 
along the direction of its buoyant trajectory
(i.e. the local radial direction in the group atmosphere).
Radiative cooling in both regions 
is driven by 
a large vortex of cluster gas flowing around the buoyant cavity 
that provides a long-lasting compression and 
differential entropy loss from the hot gas. 
Our computed 
cooling in the radial filament and cavity regions combine to 
create a filamentary $\Upsilon$-shaped cooling 
pattern in projection  
that resembles, on a smaller scale, 
the iconic post-cavity flow observed in the big buoyant 
northwestern cavity the Perseus cluster. 

As expanding X-ray cavities
form in the hot atmospheres of galaxy
groups and clusters, they drive shocks into the
hot gas that provide an important source of feedback heating
required to offset radiative losses
and catastrophic central cooling.
In this paper we show that
cavities also stimulate limited radiative cooling in
cluster gas that is compressed in the wakes of
their buoyant flow.
Although the net effect of cavities is to increase the
entropy of cluster atmospheres by shock dissipation etc.,
cavities also lower the entropy in significant and observable
masses of gas,
some or most of which may cool to form
gravitationally unbound molecular regions
similar to those observed in the NGC 5044 group
(David et al. 2014)
and elsewhere (e.g. O'Sullivan et al. 2014).

Localized cooling dropout continues long after the X-ray 
cavity disrupts and is difficult to observe.
Late time inhomogeneous cooling 
in the cavity region ($z > 1$ kpc $R > 0.3$ kpc) 
can dominate the total mass of 
gas that ultimately cools near the center.
Evidently, the range in times gas spends 
at various densities during the 
early vortex compression imprints a 
range of lowered entropy that cools 
at various times afterwards.
This later cooling occurs in isolated regions or 
in small non-radial filaments/regions distributed 
nearly isotropically near the cluster center.
More energetic cavities stimulate more off-center cooling.
The total mass of off-center (cavity region) cooled 
(dropped out) gas created by a single cavity 
of energy $10^{56} - 10^{57}$ ergs in the galaxy group NGC 5044, 
$10^8-10^9$ $M_{\odot}$, 
exceeds the total mass of extended 
molecular gas $\sim10^8$ $M_{\odot}$ recently observed 
in NGC 5044 with the 30m IRAM and ALMA telescopes.

Our cavity calculations are performed as a 
cooling flow develops in the cluster atmosphere.
The cooling flow 
is modified, but not eliminated, by a single cavity 
which cannot provide enough sustained energy. 
Cooling flows can be stopped by computing 
a series of intermittent cavities having various 
energies and distances from the cluster center (Mathews 2009).
When feedback cavity energy balances radiative losses,
the time-averaged gas density and temperature profiles
in the hot atmosphere are kept roughly constant.
When this balance is achieved,
the net radial flow of cluster gas approaches 
zero, but radial gas motions are by no means eliminated.
Even when cooling flows are largely arrested,
some buoyant outflow of high-entropy gas 
continues due to recent 
cavities.
Low entropy gas 
first flows outward in the vortex flow 
that compresses against the buoyant cavity, 
then, having radiated 
away some of its entropy, ultimately sinks back toward
the cluster center taking $\sim10^8$ yrs.
As this low-entropy, slightly denser gas 
approaches the cluster core,
it cools to low temperatures 
before arriving at the cluster center. 
Consequently,
when a cooling flow is properly arrested with 
multi-cavity feedback, 
we may expect warm and/or molecular clouds to 
form near the central galaxy.

\acknowledgements
Studies of
the evolution of hot gas in elliptical galaxies at UC Santa Cruz 
were supported during the earlier phases of this work by the NSF
and are currently supported by a NASA-funded Chandra Theory Grant for 
which we are very grateful. 
FB is supported in part by the Prin MIUR grant 2010LY5N2T``The
Chemical and
Dynamical Evolution of the Milky Way and Local Group Galaxies''.



\clearpage


\clearpage

\end{document}